# Stacking order-dependent sign-change of microwave phase due to eddy currents in nanometer-scale NiFe/Cu heterostructures

O. Gladii,[*] R. L. Seeger,[*] L. Frangou,[*] G. Forestier, U. Ebels, S. Auffret, and V. Baltz[**]

*SPINTEC, Univ. Grenoble Alpes / CNRS / INAC-CEA / GINP, F-38000 Grenoble, France*

[*] *equal contribution*

[**] *vincent.baltz@cea.fr*

**Abstract**

In the field of spintronics, ferromagnetic/non-magnetic metallic multilayers are core building blocks for emerging technologies. Resonance experiments using stripline transducers are commonly used to characterize and engineer these stacks for applications. Up to now in these experiments, the influence of eddy currents on the excitation of the dynamics of ferromagnetic magnetization below the skin-depth limit was most often neglected. Here, using a coplanar stripline transducer, we experimentally investigated the broadband ferromagnetic resonance response of NiFe/Cu bilayers a few nanometers thick in the sub-skin-depth regime. Asymmetry in the absorption spectrum gradually built up as the excitation frequency and Cu-layer thickness increased. Most significantly, the sign of the asymmetry depended on the stacking order. Experimental data were consistent with a quantitative analysis considering eddy currents generated in the Cu layers and the subsequent phaseshift of the feedback magnetic field generated by the eddy currents. These results extend our understanding of the impact of eddy currents below the microwave magnetic skin-depth and explain the lineshape asymmetry and phase lags reported in stripline experiments.

Resonance experiments are a powerful means to study physical systems and facilitate advances in material characterization and engineering. In the field of spintronics, ferromagnetic/non-magnetic (F/NM) metallic multilayers are core building blocks for emerging technologies.[1] In these multilayers, the physical properties of the F (effective magnetization, anisotropy, damping), the NM metal (spin penetration length, relaxation mechanisms, eddy currents) and the interface (spin filtering, roughness) can all be recorded by measuring ferromagnetic resonance (FMR) spectra and determining their position (resonance field), linewidth, and lineshape.[2]

Lineshape asymmetries are relatively common in FMR experiments performed with stripline setups (coplanar and microstrip).[3–7] The part of the stripline inductively coupled to the sample is equivalent to a device circuit defining a complex microwave impedance.[3,8,9] The resulting phase of the microwave excitation leads to an absorption-dispersion admixture and produces asymmetric lineshapes. Although a phenomenological parameter accounting for such asymmetry is considered to extract the resonance field and linewidth from data fitting, it is usually not commented on. The reason for this is because in most cases, asymmetry, linewidth and resonance field are not related, and because for most geometries, the absorption component prevails[3] and this type of *experiment-related asymmetry* is therefore small.

Other effects such as eddy currents may produce unusual lineshapes. This type of effect has been thoroughly studied for film thicknesses above the skin-depth limit.[10] In contrast, below this limit, the effects of eddy currents were most often neglected, except for a study on ac charge currents, including currents produced by spin pumping and spin-charge conversion[11] and for series of comprehensive studies focused on microwave screening/shielding,[3,9,12,13] e.g. leading to layer-transducer ordering-dependent standing spin wave modes in sufficiently thick layers[3,12] and to depth-dependent dephasing.[13] As we will further discuss below, eddy currents need to be carefully considered to accurately determine damping[14,15] and other related spintronic properties,[16] especially when characterizing low-damping materials. Some recent experimental studies on F-NiFe(10nm)/NM-(Au,Cu) bilayers[17,18] revealed how the Oersted field - due to eddy currents in the NM layer - affects the dynamics of F magnetization, and more specifically, how it distorts the resonance lineshape. The experiments were performed in a cavity setup and corroborated the results of analytical calculations. The scenario considered in Refs. [17,18] involved eddy currents in the NM conductor, generated by homogeneous excitation radiofrequency magnetic field ($\mathbf{h}_{rf}$) applied out-of-plane. The phaseshift ($\varphi$) between $\mathbf{h}_{rf}$ and the eddy current-induced field out-of-plane ($\mathbf{h}_{ind} = \mu \mathbf{h}_{rf} e^{i\varphi}$) resulted in an absorption($A$)-dispersion($D$) admixture of the signal. It produced an asymmetric resonance line, related to the absorbed power, $P \propto A + \beta D$ with:

$$\beta = \frac{\mu \sin(\varphi)}{1 + \mu \cos(\varphi)} + \beta_0 \qquad (1),$$

where $\beta_0$ is the empirical residual *experiment-related* phaseshift. $P$ can be calculated from: $P \propto \mathrm{Re}\left[i\omega(\chi h_{rf}) h_{rf}^*\right]$, where the magnetic susceptibility $\chi$ is deduced from the Landau-Lifshitz-Gilbert equation. In this scenario relying on the use of homogeneous $\mathbf{h}_{rf}$ out-of-plane, experiments conducted with stripline setups, with $\mathbf{h}_{rf}$ in the sample plane, should not generate eddy currents in the conductive layers. However, it has been suggested that sample tilting would lead to an out-of-plane component,[17] thus creating the conditions for *eddy current-related asymmetry*. According to this hypothesis, the sign of the asymmetry should be independent of the stacking order for the layers, because the homogeneous $\mathbf{h}_{rf}$ generates eddy currents with $\mathbf{h}_{ind}$ pointing in the same direction above and below the NM conductor. The data show that this assumption fails to completely describe experimental results.

In this article, the incompletely understood impact of eddy currents is investigated and we unravel the contributions to lineshape asymmetry in stripline experiments. The full stacks used in this study were (from substrate to surface): Cu($t_{Cu}$)/NiFe($t_{NiFe}$)/Al(2)Ox and NiFe($t_{NiFe}$)/Cu($t_{Cu}$)/Al(2)Ox (nm) multilayers. $t_{Cu}$ is the thickness of the Cu layer and was varied between 1 and 14 nm; $t_{NiFe}$ is the thickness of the NiFe layer: $t_{NiFe}$ = 4, 8, or 12 nm. Stacks were deposited at room temperature by dc-magnetron sputtering on Si/SiO$_2$(500) substrates at a pressure of argon of 2.3 x 10$^{-3}$ mbar. The NiFe layer was deposited from a Ni$_{81}$Fe$_{19}$ (at. %) permalloy target. A 2-nm-thick Al cap was deposited to form a protective Al(2)Ox film after oxidation in air. This insulating film also prevented electrical contact between the samples and the waveguide (Fig. 1(a,b)). Unless specified otherwise, the sample dimensions were: $l$ = 4 mm, $w$ = 3 mm. The microwave transducer consisted of a double-ground plane broadband coplanar waveguide. The width of the central conductor strip was 375 µm and the gap between the lines was: $g$ = 140 µm. A Schottky diode was used for detection. Modulation coils and lock-in detection were used to enhance the signal-to-noise ratio. FMR experiments and the corresponding differential absorption spectra, $d\chi''/dH \propto dP/dH$ vs. $H$ (Fig. 1(c-h)), were recorded at room temperature at frequencies ranging between 4 and 20 GHz.

First, we discuss the spectrum asymmetry which gradually built up as the $t_{Cu}$ increased (Fig. 1(c-h)). This behavior revealed a non-negligible impact of eddy currents circulating in the Cu layers. Most importantly, the sign of the asymmetry depended on the ordering of the Cu and NiFe layers, i.e., whether the Cu layer was the buffer or capping layer. Figure 2(a) illustrates a mechanism where the inhomogeneous field of the coplanar waveguide, with out-of-plane components, generates eddy currents with oppositely directed induced in-plane fields ($\mathbf{h}_{ind}$) above and below a Cu layer. Further considering a phaseshift between $\mathbf{h}_{rf}$ and $\mathbf{h}_{ind}$, one can write: $\mathbf{h}_{ind}^{\pm} = \mu \mathbf{h}_{rf} e^{i\varphi^{\pm}}$. The superscripts '+' and '-' relate to the 'capping' and 'buffer' cases respectively. We get a situation similar to Ref. 17, described by Eq. (1), but considering that the dominant coupling occurs between the FMR mode and the in-plane fields. To extract the asymmetry and quantify the findings, the differential resonance spectra were fitted using the following equation:[17] $\frac{d\chi''}{dH} \propto \frac{d}{dH}\left[\frac{1+\beta(H-H_{res})/(\sqrt{3}\Delta H_{pp})}{(H-H_{res})^2+(\sqrt{3}\Delta H_{pp}/2)^2}\right]$, where $H_{res}$ is the resonance field, $\Delta H_{pp}$ is the peak-to-peak linewidth.

Figure 2(b) shows $\beta$ plotted as a function of $t_{Cu}$ for series of Cu($t_{Cu}$)/NiFe($t_{NiFe}$=4;8;12)/Al(2)Ox ('buffer') and NiFe($t_{NiFe}$=4;8;12)/Cu($t_{Cu}$)/Al(2)Ox (nm) ('capping') multilayers. The gradual increase in $|\beta|$ with $t_{Cu}$ agrees with the fact that eddy currents relate to the conductance of the Cu layers, which increases with $t_{Cu}$. The above deductions can be correlated by using Eq. (1). The field $h_{ind}$ relates to the rate of change of magnetic flux through the area, $S$, delimited by the eddy current loop. It can be expressed as $h_{ind} = \mu_0 2\pi f \frac{t_{Cu}}{\rho} a(l,w,g) h_{rf}$. We considered that the eddy current was given by $I = S 2\pi f h_{rf}/R$, where the numerator corresponds to the electromotive force due to variations of $\mathbf{h}_{rf}$ over time. The resistance of the loop of length $P$, is given by $R = \rho P/(t_{Cu}\zeta)$, where $\zeta$ is the width of the loop, the spatial profile of which depends on the sample's geometry in a non-trivial manner.[19] The averaged magnetic field acting on the NiFe layer is expressed as $h_{ind} = \mu_0 I/b$, where $b$ is a function of the geometry of the sample. The geometry-dependence of the parameters, including $S\zeta/(bP) \equiv a(l,w,g)$ (dimension of length) will be discussed further below. Over the thickness range investigated (1-14 nm), the Cu layer's resistivity is given by $\rho_{Cu}^{\pm} = \rho_0^{\pm} + \eta^{\pm}/t_{Cu}$, where $\eta^{\pm} = 3\lambda_{mfp}^{\pm}/8$ according to the Fuchs-Sondheimer model.[20,21] From separate 4-point electrical measurements, we obtained $\rho_{Cu}^+[\mu\Omega.cm] = 7 + 53/t_{Cu}[nm]$ for the capping layers and $\rho_{Cu}^-[\mu\Omega.cm] = 5 + 35/t_{Cu}[nm]$ for the buffer layers (Fig. 2(c)), which produce reasonable values for the electron mean free path: $\lambda_{mfp}^+ = 14nm$ and $\lambda_{mfp}^- = 9nm$. With regard to the phaseshift ($\varphi$) in Eq. (1), we neglected the contribution of the skin effect, which is proportional to $t_{Cu}/\delta_{Cu}$,[22] because $t_{Cu}$=1-14 nm, and the skin-depth $\delta_{Cu}$~1000-500 nm for f=4-20 GHz. In the ideal situation of a negligible inductive contribution to the complex impedance, $\varphi^+ = \pi/2$ for the 'capping' layer case (quadrature phaseshift because $\mathbf{h}_{ind}$ is related to the time derivative of $\mathbf{h}_{rf}$), and $\varphi^- = -\pi/2$ for the 'buffer' layer case (in antiphase to the 'capping' layer case). Developing the different terms in Eq. (1) produced a predictable non-linear dependence of $\beta$ on $t_{Cu}$, and a linear dependence on $f$:

$$\beta^\pm = \pm\mu_0 2\pi f \frac{t_{Cu}}{\rho_0^\pm + \eta^\pm/t_{Cu}} a(l,w,g) + \beta_0 \qquad (2).$$

The two solid lines in Fig. 2(b) were fitted with Eq. (2); $a(l,w,g)$ and $\beta_0$ were the only free parameters. It can be seen that the simplified model captures the physics of the phenomenon observed experimentally. Data fitting returned $\beta_0$=-0.3 and $a(l,w,g)$=185±3 µm in both cases, in agreement with the constant sample dimensions in these sets of experiments. Remarkably, the model can account for the difference in the thickness-dependence of Cu-resistivity due to the inversion of the growth order. To emphasize this, the dashed line in Fig. 2(b) corresponds to a simulation using Eq. (2), considering the fictitious case of $\rho_{Cu}^+ = \rho_{Cu}^-$. From Fig. 2(b), we note that slight deviations between predictions and experimental data can still be observed for the 'buffer' layer case. Most importantly for thick Cu layers - as for example experimentally shown in Ref.[17] in the 10-50 nm range - and high frequencies, inductive contributions to the complex impedance are very likely to affect the ideal thickness-dependence of $\beta$ in a non-trivial manner. Considering such a term, the phase shift becomes: $\varphi = \pm\pi/2 + \theta(f, t_{Cu}, l, w, g)$, where $\theta$ shows a non-linear dependence on several parameters, thus producing non-linear dependences of $\beta$ (from Eq. (1)).

The non-trivial influence of the inductive contributions to the complex impedance can clearly be seen for $f$-dependent measurements. Figure 2(d) shows $\beta$ vs. $f$, for series of Cu($t_{Cu}$=1;8;14)/NiFe(8)/Al(2)Ox 'buffer' and NiFe(8)/Cu($t_{Cu}$=1;8;14)/Al(2)Ox (nm) 'capping' multilayers. Data for $t_{Cu}=1$ nm, in the absence of eddy current, correspond to $\beta_0$ and superimpose for the 'buffer' and 'capping' cases. The *f*-dependence of $\beta_0$ is weak, ruling out any *f*-dependent impedance contribution of the NiFe layer to the phaseshift. The solid lines in Fig. 2(d) were produced by calculations using Eq. (2). The same set of parameters as that returned from Fig. 2(b) was used. It concurrently described the thickness- and *f*-dependences of $\beta$ for the 'capping' case (Fig. 2(d)), confirming that the simplified model reflects the physics behind the phenomenon observed. The overall linear increase of $|\beta|$ with *f*, driven by the fact that eddy currents increase when the rate of change of flux rises, may be altered by complex inductive contributions, which are known to increase for higher frequencies and thicker films. In agreement with this information, we observe in Fig. 2(d) that data depart from a linear dependence above 10 GHz for the 14-nm-thick layers, a result that contrasts with those obtained for the 8-nm-thick ones, which follow a linear dependence throughout. The 14-nm-thick buffer layer case typically illustrates how non-trivial contributions can drastically distort and bend the initially linear *f*-dependence. To rule out any contribution of the Si/SiO$_2$(500) substrate on the sign-change of $\beta$, we compared a Cu(14)/NiFe(4)/Al(2)Ox to a NiFe(4)/Cu(14)/Al(2)Ox (nm) stack deposited on glass substrates (not shown). The same trend of a positive vs. negative value of $\beta$ for the 'capping' vs. 'buffer' case was obtained.

We will now consider finite-size effects. Once again using Fig. 2(b), we will briefly comment on the square crossed symbol corresponding to data recorded after patterning only the Cu(14)/Al(2)Ox capping layers in a Si/SiO$_2$/NiFe(8)/Cu(14)/Al(2)Ox (nm) stack (inset of Fig. 2(a)). A 4x3 mm$^2$ array of square dots with lateral size of 100 µm was fabricated. Following patterning, two effects compete with one another. First, the number of eddy current loops increases, and simultaneously, the path of each loop is constrained. The fact that patterning reduced $\beta$ to a value close to $\beta_0$ (Fig. 2(b)) shows that eddy currents cannot develop in the dots. This result indicates that the dot size was smaller than the width of the eddy current loop. We further assessed the dependence of $\beta$ on the sample's geometry in Fig. 3(a). We considered geometry-dependent parameters, $S\zeta/(bP) \equiv a(l,w,g)$ to account for the fact that the spatial profile of the eddy currents depends on the sample's geometry in a non-trivial manner. In particular, the width ($\zeta$) and the circulation (determining $S$ and $P$) are unknown. In addition, the amplitude of $I$ is likely inhomogeneous along the width $\zeta$, making it difficult to obtain an analytical expression for the parameter $b$ relating to the magnetic field created by $I$. Considering the limit case when eddy currents extend over the full sample (inset in Fig. 3(a)), we obtained a linear dependence on $S/P$ (Fig. 3(a)), meaning that $\zeta/b$ seems to be almost independent of the geometry. The discrepancy for the $\beta_0$ intercept is probably related to a geometry-dependence close to the smallest dimensions, that is likely to result in curve-binding. The results also show that rotating the sample in the plane of the stripline had no impact on the data (Fig. 3(a)), demonstrating that both the length and the width of the current path contribute to $h_{ind}$. The most relevant insight is that stacking order-, thickness-, and f-dependent measurements (Fig. 2) returned the same value of $a(l,w,g)$, in agreement with the constant sample dimensions in these sets of experiments. The value of $a$~185 µm is also in agreement with the order of magnitude that can be estimated from Ref. 17. Figures 3(b,c) present control data showing that the sample vs. stripline dimensions remained within a range where the linewidth ($\Delta H_{pp}$) and resonance field ($H_{res}$) were unaffected by geometrical effects.

Before concluding the paper, we will briefly comment on $H_{res}$ and $\Delta H_{pp}$. The total Gilbert damping, $\alpha$, was calculated from the slope of f-dependent measurements ($\Delta H_{pp}$ vs. $f$), from $\Delta H_{pp} = \Delta H_{pp0} + 4\pi\alpha f/(\sqrt{3}|\gamma|)$,[23] where $\Delta H_{pp0}$ is the inhomogeneous broadening[24] due to spatial variations in the magnetic properties (values of a few Oe were measured in our experiments) and $\gamma$ is the gyromagnetic ratio (derived from the fit of the curve representing $H_{res}$ vs. $f$ ). Plots representing $H_{res}$ vs. $t_{Cu}$ and $\alpha$ vs. $t_{Cu}$ are shown in Figs. 4(a,b) and Figs. 4(c,d), respectively. The data showed no obvious link between eddy currents in the Cu layers (spectrum asymmetry in Fig. 2), and the spectrum position, $H_{res}$. Regarding $\alpha$, eddy currents in conductors adjacent to a resonator, including the waveguide, were shown to contribute to a damping process due to losses via inductive coupling.[14,15,25] This phenomenon is referred to as radiation damping and can be expressed as $\alpha^{rad} = \mu_0^2 \kappa \gamma M_S t_F w / (2Z_0 l)$,[14] where $M_S$ is the saturation magnetization, $t_F$ is the ferromagnet's thickness, $Z_0$ is the impedance of the NM conductor, and $\kappa$ accounts for the mode profile. For YIG(200)/Al2O3(30)/Pt(5-20) (nm) samples with $M_S \sim$ 121 emu.cm$^{-3}$ and dimensions of 2x5 mm$^2$, in-plane stripline FMR measurements showed that $\alpha^{rad}$ due to eddy currents in the Pt capping layer can be up to $3\times10^{-4}$ - for the 20 nm thick 35-$\Omega$ Pt layer.[15] From this value, considering the dependence of $\alpha^{rad}$ on $M_S$, $t_F$, $Z_0$ and the sample dimensions, and extrapolating to our case, we estimate a maximum value of $\alpha^{rad}$ of $1\times10^{-4}$ for the NiFe(12)/Cu(14) (nm) with $M_S \sim$ 700 emu.cm$^{-3}$, dimensions of 3x4 mm$^2$ and a resistance of the Cu layer of 10 $\Omega$. This value of $\alpha^{rad}$ is too small to influence the damping of our NiFe layers ($\alpha \sim 6$-$8\times10^{-3}$). In fact, no obvious contribution of eddy currents to $\alpha$ can be inferred from our data. However, given the orders of magnitude indicated above, radiation contribution due to eddy currents in NM layers will need to be carefully considered when extracting $\alpha$ in several other cases. For example, with a 5-fold increase of the NiFe layer thickness, $\alpha^{rad}$ will become sizeable. In addition, an $\alpha^{rad}$ of the order of few $10^{-4}$ is already significant for materials exhibiting low intrinsic damping, such as the YIG insulator ($\alpha \sim 6\times10^{-5}$),[15] the $Co_{1.9}Mn_{1.1}Si$ half metal Heusler alloy ($7\times10^{-4}$),[26] and the $Co_{25}Fe_{75}$ bcc alloy ($5\times10^{-4}$).[27] We finally note from Fig. 4 that, for $t_{NiFe} = 4$ nm, a non-monotonous dependence of $H_{res}$ was observed. This behavior supports non-monotonous dependence of the effective NiFe magnetization, $M_{eff}$, (Figs. 4(e,f)) which can be extracted from $H_{res}$ vs. $f$ using the Kittel formula:[28]

$(2\pi f)^2 = |\gamma| H_{res}(H_{res} + 4\pi M_{eff})$. We recall that $M_{eff} = M_S - 2K_S/(4\pi M_S t_{NiFe})$. The values of $M_S$ (Fig. 4(g,h)), measured independently by magnetometry, were monotonous and thus confirmed that the non-monotonous behavior of $M_{eff}$ seems to primarily relate to the properties of the Cu/NiFe interface. A similar non-monotonous dependence of α was observed. Cu wets poorly on SiO$_2$ compared to NiFe on SiO$_2$ and Cu, and as a result may create rougher thin Cu films. Consequently, spatially inhomogeneous stray fields may lead to incoherent dephasing of the spin current[29,30] injected from the NiFe to the buffer Cu layer, and thus to enhanced damping.

In conclusion, the main contribution of this paper is that it represents systematic experimental evidence of a stacking-order-dependent sign-change of the microwave phase in nanometer-scale NiFe/Cu bilayers. The effect could be ascribed to eddy currents generated in the Cu layer in the sub-skin-depth regime by the time varying magnetic fields in the experiment. Distinct sets of experimental data were consistent with a simple quantitative analysis encompassing the main features of the phenomenon. These results contribute to our understanding of the impact of eddy currents below the microwave magnetic skin-depth and explain the contributions to lineshape asymmetry and phase lags reported in stripline experiments commonly used to characterize and engineer materials for spintronic applications. They support a rational explanation to the use of the 'phenomenological' parameter accounting for lineshape asymmetry when extracting the spectral resonance field and linewidth from FMR data-fitting. The results also provide a straightforward way to detect the contributions of eddy currents from NM-adjacent conductors, as a caveat for the need in some cases to take these contributions into account when attempting to accurately determine damping[14,15] and other related spintronic properties such as spin-mixing conductance and the spin-Hall angle in spin-pumping experiments.[16]

**Acknowledgments**

We acknowledge financial support from the French national research agency (ANR) [Grant Number ANR-15-CE24-0015-01] and KAUST [Grant Number OSR-2015-CRG4-2626]. We also thank M. Gallagher-Gambarelli for critical reading of the manuscript.

**Figure captions**

Fig. 1. (color online) (a,b) Schematic representations of the coplanar waveguide (CPW) – FMR experiment. Samples were placed face-down on the waveguide. The in-plane dc bias field (**H**), and the in-plane component of the excitation magnetic field from the waveguide (**h**$_{rf}$) are represented. (c-h) Representative differential absorption spectra ($d\chi''/dH$ vs. $H$) measured for Si/SiO$_2$/NiFe(8)/Cu(t$_{Cu}$)/Al(2)Ox and Si/SiO$_2$/Cu(t$_{Cu}$)/NiFe(8)/Al(2)Ox (nm) stacks. The solid lines were fitted to the data using a model derived from Ref. [17], which is described in the text. Data-fitting allowed the resonance field ($H_{res}$), the peak-to-peak line width ($\Delta H_{pp}$), and the asymmetry parameter ($\beta$) to be determined.

Fig. 2. (color online) (a) Illustration of a mechanism where the inhomogeneous field (**h**$_{rf}$) of the coplanar waveguide, with strong out-of-plane components, generates eddy currents with oppositely directed induced in-plane fields (**h**$_{ind}$) above and below a Cu layer. (b) Representative series of dependences of $\beta$ on 'capping' and 'buffer' Cu-layer thickness (t$_{Cu}$) for Si/SiO$_2$/Cu(t$_{Cu}$)/NiFe(t$_{NiFe}$=4;8;12)/Al(2)Ox and Si/SiO$_2$/NiFe(t$_{NiFe}$=4;8;12)/Cu(t$_{Cu}$)/Al(2)Ox (nm) stacks. Data were recorded at 10 GHz. The square crossed symbol corresponds to data recorded after patterning (inset) the Cu(14)/Al(2)Ox bilayer in a Si/SiO$_2$/NiFe(8)/Cu(14)/Al(2)Ox (nm) stack. The open square symbol corresponds to data for the same stack on which the whole etching process was performed, as these samples were protected by a resist they remained unpatterned. (c) Corresponding dependences of Cu-layer resistivity ($\rho_{Cu}$) on t$_{Cu}$, obtained separately using standard 4-point electrical measurements. The lines were obtained using linear fits. (d) Representative series of dependences of $\beta$ on frequency ($f$) for Si/SiO$_2$/Cu(t$_{Cu}$=1;8;14)/NiFe(8)/Al(2)Ox and Si/SiO$_2$/NiFe(8)/Cu(t$_{Cu}$=1;8;14)/Al(2)Ox (nm) stacks. The solid lines in (b) and (c) were obtained using the model described in the text. The dashed and dash-dotted lines are visual guides.

Fig. 3. (color online) (a) Dependences of $\beta$ on the ratio *S/P* for Si/SiO$_2$/NiFe(8)/Cu(14)/Al(2)Ox (nm) stacks for two sample's orientations: *l // h$_{rf}$* and *w // h$_{rf}$*. *S* is the area delimited by the eddy current loop and *P* is the length of the current path, considering that currents extend over the sample (see inset). For *l // h$_{rf}$ S/P=(wl/2)/[2(w+l/2)]*, and for *w // h$_{rf}$ S/P=(lw/2)/[2(l+w/2)]*. Data were recorded at 10 GHz. The line in (a) is a visual guide. (b,c) Corresponding dependences of $H_{res}$ and $\Delta H_{pp}$.

Fig. 4. (color online) Dependences of (a,b) $H_{res}$, (c,d) $\alpha$, (e,f) $M_{eff}$, and (g,h) $M_S$ on t$_{Cu}$ for Si/SiO$_2$/Cu(t$_{Cu}$)/NiFe(t$_{NiFe}$=4;8;12)/Al(2)Ox and Si/SiO$_2$/NiFe(t$_{NiFe}$=4;8;12)/Cu(t$_{Cu}$)/Al(2)Ox (nm) stacks. The square crossed symbols correspond to the patterned sample. (a-b) correspond to data recorded at 10 GHz. (c-d) were deduced from *f*-dependences of $\Delta H_{pp}$. (e,f) were deduced from *f*-dependences of $H_{res}$. (g,h) were measured independently by magnetometry, using a superconducting quantum interference device.

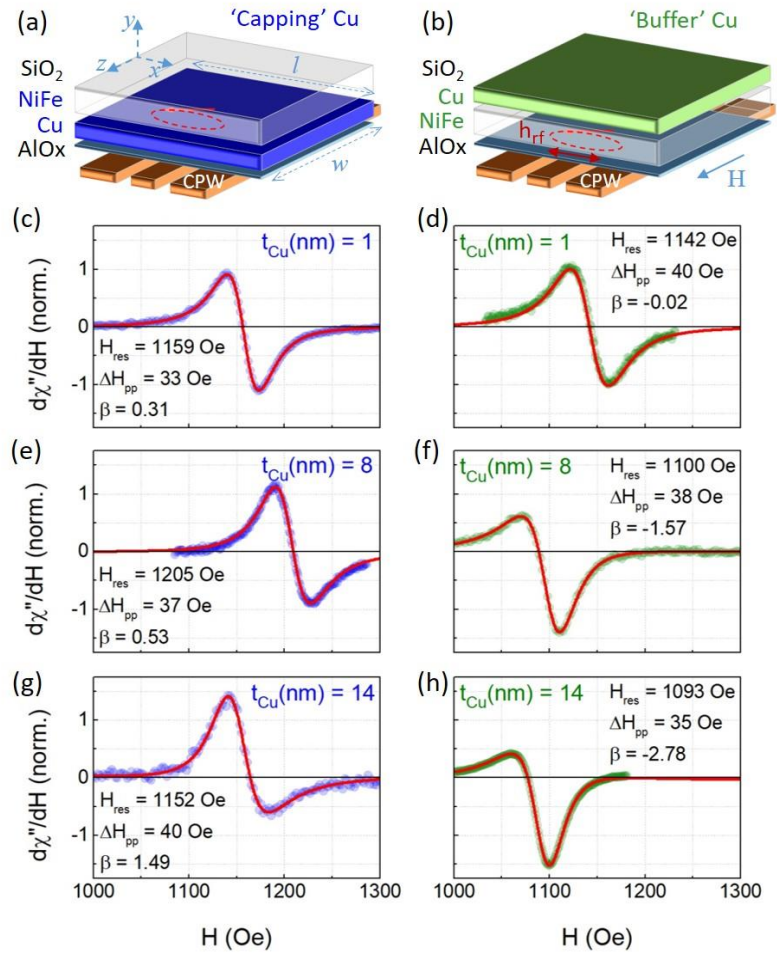

Fig. 1

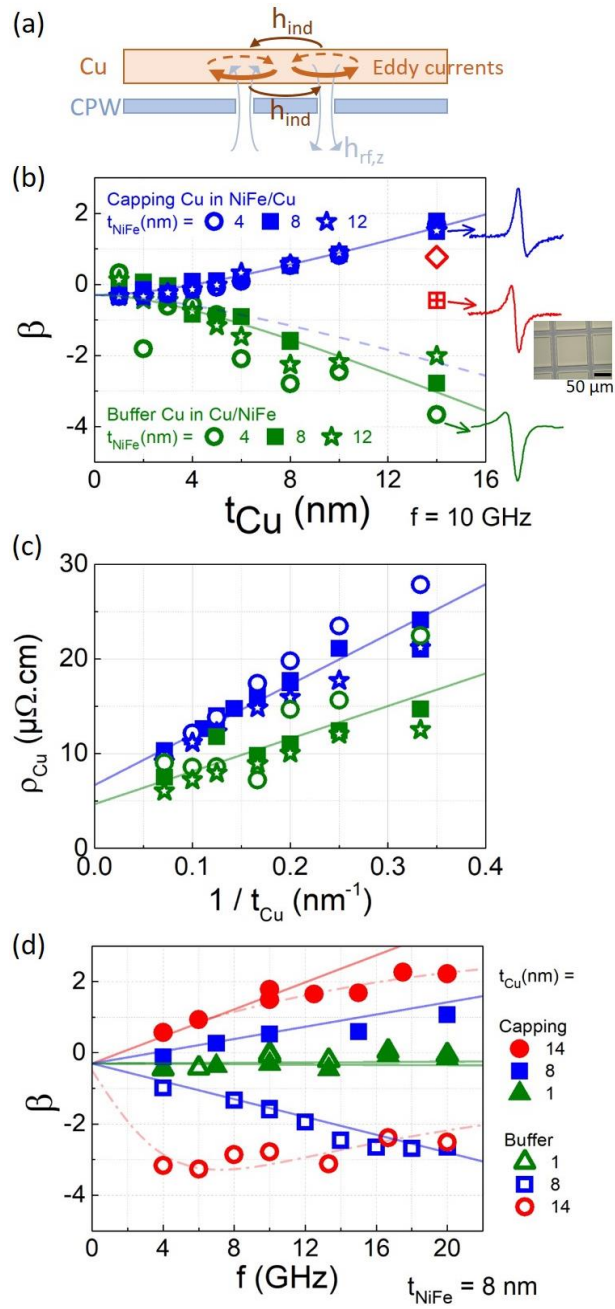

Fig. 2

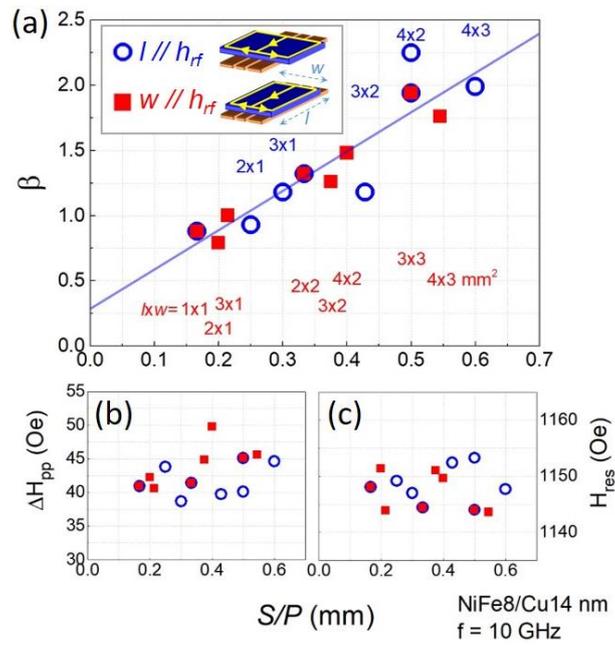

Fig. 3

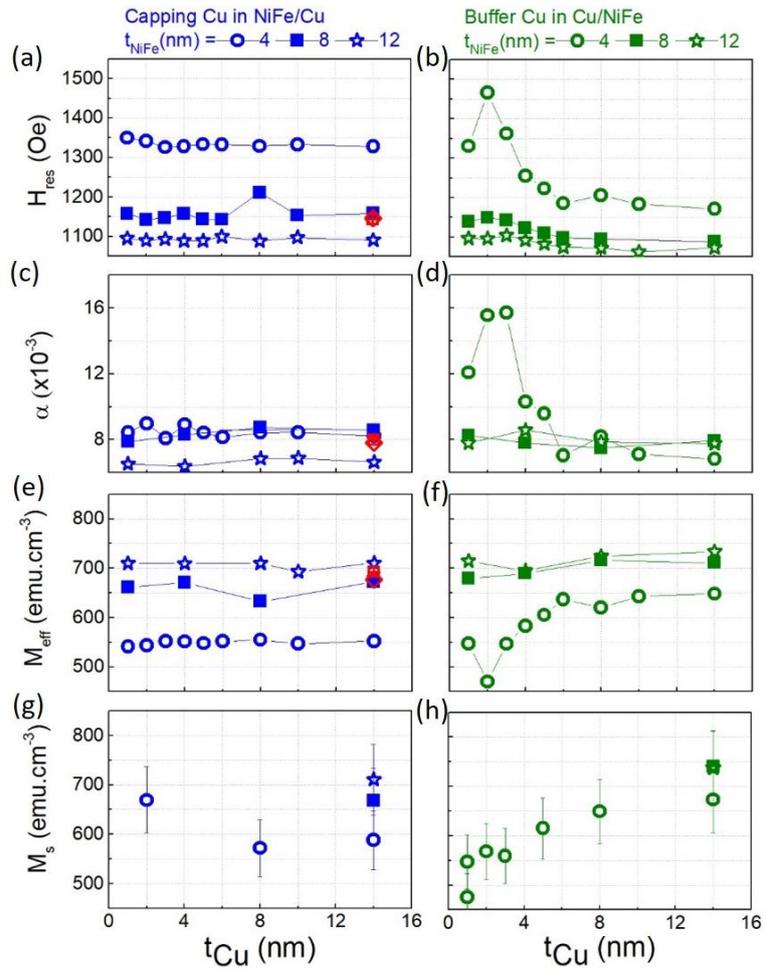

Fig. 4